\documentclass[a4paper,11pt]{article}

\usepackage[T1]{fontenc}   
\usepackage[utf8]{inputenc} 
\usepackage{lmodern}        
\usepackage{upquote}        
\usepackage{listings}
\usepackage{amsmath,amssymb,mathtools,bm}
\numberwithin{equation}{section}
\allowdisplaybreaks
\DeclareMathOperator{\atanTwo}{atan2}

\usepackage{graphicx}
\usepackage{tikz}
\usetikzlibrary{arrows.meta,calc}
\tikzset{>=Stealth}
\usepackage{float}
\usepackage{comment}
\usepackage[a4paper,margin=28mm]{geometry}
\usepackage{microtype}
\usepackage{xcolor}
\usepackage[font=small,labelfont=bf]{caption}

\usepackage{listings}
\lstdefinestyle{code}{
  basicstyle=\ttfamily\footnotesize,
  numbers=left, numberstyle=\tiny, stepnumber=1,
  frame=single, framerule=0.3pt,
  breaklines=true, breakatwhitespace=true,
  columns=fullflexible, showstringspaces=false,
  upquote=true,
  keywordstyle=\bfseries,
  commentstyle=\itshape,
}

\usepackage{hyperref}
\hypersetup{
  colorlinks=true,
  linkcolor=blue!50!black,
  citecolor=blue!50!black,
  urlcolor =blue!50!black
}

\title{Analysis of Three-Particle Elastic Collisions Using Newtonian\\
       Mechanics and Vector Geometry}
\author{Shuhei Kobayashi%
\thanks{Corresponding author: \texttt{xiaolinxiuping1@gmail.com}}}

\date{August 31, 2025} 
\begin{document}
\maketitle
\begin{center}
Department of Physics and Astronomy, Faculty of Science and Technology, Tokyo University of Science, Noda City, Chiba, Japan\\
\texttt{6222041@ed.tus.ac.jp} \quad (long-term: \texttt{xiaolinxiuping1@gmail.com})
\end{center}
\begin{abstract}
We study one-dimensional elastic collisions of three point masses on a line under vacuum, with no triple collisions. We express momentum conservation in matrix form and analyze the composite map
$D=D_{BC}D_{AB}$ and its powers $D^k$, which yield the velocities after any prescribed
number of collisions for arbitrary mass ratios and initial data.
 After that, using vector $\bm{u}$ on a plane $s^\perp$, the total number of collisions is
\[
n\;=\;1+\Big\lfloor\tfrac{\Omega-\phi_{BC}}{\theta}\Big\rfloor+\Big\lceil\tfrac{\Omega-\phi_{BC}}{\theta}\Big\rceil,
\] Through this concept, $D$ is recognised as giving $\bm{u}$ a rotation with angle $\theta$ which is determined by only mass ratios. And, we calculated energy transfer through collisions. With the work, we find that the change of energy is proportional to total momentum of two particles and average velocity of particles based on initial average velocity of A and B before collision.
\end{abstract}

\section{Setting}
Assume three point masses A,B,C on the $x$-axis. Let their masses be $m_A,m_B,m_C$, with initial velocities $v_{A0},v_{B0},v_{C0}$ and initial positions $x_A>x_B>x_C$ (no overtaking before the first collision). The system is in vacuum; no forces act except impulsive forces at impacts. Triple collisions do not occur, and all collisions are elastic.
 When there is a collision between A and B, velocities of each particle after the collision $v_A'$ and $v_B'$ are
\begin{align}
v_A'=\frac{m_A-m_B}{M_{AB}}v_A+\frac{2m_B}{M_{AB}}v_B\\
v_B'=\frac{2m_A}{M_{AB}}v_A+\frac{m_B-m_A}{M_{AB}}v_B
\end{align}
where $M_{AB}=m_A+m_B$. Similarly, after the collision between B and C, their velocities are
\begin{align}
v_B'=\frac{m_B-m_C}{M_{BC}}v_B+\frac{2m_C}{M_{BC}}v_C\\
v_C'=\frac{2m_B}{M_{BC}}v_B+\frac{m_C-m_B}{M_{BC}}v_C
\end{align}
where $M_{BC}=m_B+m_C$. In matrix form, these become:
\begin{align}
\mathrm{AB}
\begin{bmatrix}
v_A'\\
v_B'\\
v_C'
\end{bmatrix}
=\frac{1}{M_{AB}}
\begin{bmatrix}
{m_A-m_B}&{2m_B}&0\\
{2m_A}&{m_B-m_A}&0\\
0&0&M_{AB}
\end{bmatrix}
\begin{bmatrix}
v_A\\
v_B\\
v_C
\end{bmatrix}
\equiv D_{AB}
\begin{bmatrix}
v_A\\
v_B\\
v_C
\end{bmatrix}
\\
\mathrm{BC}
\begin{bmatrix}
v_A'\\
v_B'\\
v_C'
\end{bmatrix}
=\frac{1}{M_{BC}}
\begin{bmatrix}
M_{BC}&0&0\\
0&m_B-m_C&2m_C\\
0&2m_B&m_C-m_B
\end{bmatrix}
\begin{bmatrix}
v_A\\
v_B\\
v_C
\end{bmatrix}
\equiv D_{BC}
\begin{bmatrix}
v_A\\
v_B\\
v_C
\end{bmatrix}
\end{align}
We enumerate collisions by an integer $n\in\mathbb{N}$ ($n=1$ for the first $AB$ impact, 
$n=2$ for the subsequent $BC$ impact, and so on). Writing $n=2k$ for even and $n=2k+1$ 
for odd, the post-collision velocities are
\[
\begin{aligned}
\text{if }n=2k:&\quad 
\begin{bmatrix}v_{A2k}\\ v_{B2k}\\ v_{C2k}\end{bmatrix}
=(D_{BC}D_{AB})^{k}\begin{bmatrix}v_{A0}\\ v_{B0}\\ v_{C0}\end{bmatrix},\\
\text{if }n=2k+1:&\quad 
\begin{bmatrix}v_{A2k+1}\\ v_{B2k+1}\\ v_{C2k+1}\end{bmatrix}
=D_{AB}(D_{BC}D_{AB})^{k}\begin{bmatrix}v_{A0}\\ v_{B0}\\ v_{C0}\end{bmatrix}.
\end{aligned}
\]
Now, first collision occurs with A and B and if the system continues to collide, next is B and C, next is A and B..., that means B collides with A and C alternately. Thus, velocities of  each particle after $n=2k$-th collision for B are 
\begin{align}
\begin{bmatrix}
v_{A2k}\\
v_{B2k}\\
v_{C2k}
\end{bmatrix}
&=(D_{BC}D_{AB})^k
\begin{bmatrix}
v_{A0}\\
v_{B0}\\
v_{C0}
\end{bmatrix}\\
&=\left(\frac{1}{M_{AB}M_{BC}}
\begin{bmatrix}
M_{BC}(m_A-m_B)&2M_{BC}m_B&0\\
2m_A(m_B-m_C)&(m_B-m_A)(m_B-m_C)&2m_CM_{AB}\\
4m_Am_B&2m_B(m_B-m_A)&M_{AB}(m_C-m_B)
\end{bmatrix}\right)^k
\begin{bmatrix}
v_{A0}\\
v_{B0}\\
v_{C0}
\end{bmatrix}
\end{align}
where $k$ is a non-negative integer.

Hereafter we write $D:=D_{BC}D_{AB}$. When $n=2k+1$, the state is 
$D_{AB}D^{k}v_0$. Thus computing $D^{k}$ suffices to obtain the velocities after any $n$.
\section{Matrix Powers}

To calculate $D^k$, SymPy calculated eigenvalues of $D$. The results $\lambda_1,\lambda_+,\lambda_-$ are below;
\begin{align}
\lambda_1&=1\\
\lambda_+&=\frac{m_Am_C-m_BM}{M_{AB}M_{BC}}+\frac{2i\sqrt{m_Am_Bm_CM}}{M_{AB}M_{BC}}\\
\lambda_-&=\frac{m_Am_C-m_BM}{M_{AB}M_{BC}}-\frac{2i\sqrt{m_Am_Bm_CM}}{M_{AB}M_{BC}}
\end{align}
where $M=m_A+m_B+m_C$. Hence $\lambda_+$ and $\lambda_-$ are complex conjugates, we shall express $\lambda_+$ as
\begin{align}
\lambda_+\equiv\gamma+i\delta\equiv\beta
\end{align}
with real number $\gamma,\delta$. At the same time, we set
\begin{align}
\theta\equiv\arg{\lambda_+}
\end{align}
And eigenvectors corresponding each eigenvalue $\bm{v}_1,\bm{v}_+,\bm{v}_-$ are
\begin{align}
\bm{v}_1&=\left(\begin{matrix}
1\\
1\\
1
\end{matrix}\right)\notag\\
\bm{v}_+&=\left(\begin{matrix}
\frac{-2m_Am_C+i\sqrt{m_Am_Bm_CM}}{m_AM_{AB}}\\
\frac{2m_Am_C-i\sqrt{m_Am_Bm_CM}}{m_AM_{AB}}\\
1
\end{matrix}\right)\notag\\
\bm{v}_-&=\left(\begin{matrix}
\frac{-2m_Am_C-i\sqrt{m_Am_Bm_CM}}{m_AM_{AB}}\\
\frac{2m_Am_C+i\sqrt{m_Am_Bm_CM}}{m_AM_{AB}}\\
1
\end{matrix}\right)
\end{align}
For simplification, we set
\begin{align}
\frac{2m_Am_C+i\sqrt{m_Am_Bm_CM}}{m_AM_{AB}}\equiv c+di
\end{align}
then
\begin{align}
\bm{v}_1&=\left(\begin{matrix}
1\\
1\\
1
\end{matrix}\right)\notag\\
\bm{v}_+&=\left(\begin{matrix}
-c+di\\
c-di\\
1
\end{matrix}\right)\notag\\
\bm{v}_-&=\left(\begin{matrix}
-c-di\\
c+di\\
1
\end{matrix}\right)
\end{align}

Therefore, with a matrix aligned with these $P=(\bm{v}_1,\bm{v}_+,\bm{v}_-)$, 
\begin{align}
D^k=P\begin{bmatrix}
\lambda_1^k&0&0\\
0&\lambda_+^k&0\\
0&0&\lambda_-^k
\end{bmatrix}
P^{-1}
\end{align}
Because $\lambda_1=1$ and $\lambda_+=\beta$, this can be rewritten as
\begin{align}
D^k=P\begin{bmatrix}
1&0&0\\
0&\beta^k&0\\
0&0&\bar{\beta}^k
\end{bmatrix}
P^{-1}
\end{align}
Now, calculating $P^{-1}$, the result is
\begin{align}
P^{-1}=\left(\frac{1}{\mathrm{det}{P}}\mathrm{adj}{P}\right)
\end{align}
. Here $\mathrm{det}{P}$ is a determinant of $P$ and $\mathrm{adj}{P}$ is a adjugate matrix. And,
\begin{align}
\mathrm{det}P&=-\frac{1}{4di}\\
\mathrm{adj}P&=\begin{bmatrix}
-2di&-2di&0\\
c+di-1&c+di+1&-2(c+di)\\
1-(c-di)&-c+di-1&2(c-di)
\end{bmatrix}
\end{align}
thus
\begin{align}
P^{-1}&=\frac{1}{4di}\begin{bmatrix}
2di&2di&0\\
1+c+di&-1-c-di&2(c+di)\\
-1+c-di&1+c-di&-2(c-di)
\end{bmatrix}\\
\therefore D^k&=\frac{1}{4di}\begin{bmatrix}
1&-c+di&-c-di\\
1&c-di&c+di\\
1&1&1
\end{bmatrix}
\begin{bmatrix}
1&0&0\\
0&\beta^k&0\\
0&0&\bar{\beta}^k
\end{bmatrix}
\begin{bmatrix}
2di&2di&0\\
1+c+di&-1-c-di&2(c+di)\\
-1+c-di&1+c-di&-2(c-di)
\end{bmatrix}
\end{align}
is concluded. Here, 
\begin{align}
|\beta|^2=\gamma^2+\delta^2=1\\
\therefore |\beta|=1>0
\end{align}
so, as
\begin{align}
\beta^k+\bar{\beta}^k=2\cos{k\theta}\\
\beta^k-\bar{\beta}^k=2i\sin{k\theta}
\end{align}
, calculating $D^k$ transforming $\beta$ into polar form to simplify the calculation, 
\begin{align*}
D^k = \frac{1}{4di} \left\{ 2i \cdot
\resizebox{\textwidth}{!}{$
\begin{bmatrix}
(c^2+d^2-c)\sin{k\theta}+d\cos{k\theta} & (c^2+d^2+c)\sin{k\theta}-d\cos{k\theta} & -2(c^2+d^2)\sin{k\theta} \\
-(c^2+d^2-c)\sin{k\theta}-d\cos{k\theta} & -(c^2+d^2+c)\sin{k\theta}+d\cos{k\theta} & 2(c^2+d^2)\sin{k\theta} \\
(1-c)\sin{k\theta}-d\cos{k\theta} & -(c+1)\sin{k\theta}-d\cos{k\theta} & 2(c\sin{k\theta}+d\cos{k\theta})
\end{bmatrix}
+ 2di
\begin{bmatrix}
1 & 1 & 0 \\
1 & 1 & 0 \\
1 & 1 & 0
\end{bmatrix}
$}
\right\}
\end{align*}
for the ease, I set
\begin{align}
c^2+d^2=\nu^2\\
\frac{v_{A0}+v_{B0}}{2}=\overline{v_{AB}}
\end{align}
and apply this to initial velocity vector which is shown as
\begin{align}
&\begin{bmatrix}
v_{A2k}\\
v_{B2k}\\
v_{C2k}
\end{bmatrix}\\
&=\frac{1}{d}
\resizebox{\textwidth}{!}{$
\begin{bmatrix}
\left[\left(\overline{v_{AB}}-v_{C0}\right)\nu^2+c\frac{v_{B0}-v_{A0}}{2}\right]\sin{k\theta}-d\frac{v_{B0}-v_{A0}}{2}\cos{k\theta}\\
-\left\{\left[\left(\overline{v_{AB}}-v_{C0}\right)\nu^2+c\frac{v_{B0}-v_{A0}}{2}\right]\sin{k\theta}-d\frac{v_{B0}-v_{A0}}{2}\cos{k\theta}\right\}\\
\left[\frac{v_{A0}-v_{B0}}{2}-c\left(\overline{v_{AB}}-v_{C0}\right)\right]\sin{k\theta}-d\left(\overline{v_{AB}}-v_{C0}\right)\cos{k\theta}
\end{bmatrix}
+\overline{v_{AB}}
\begin{bmatrix}
1\\
1\\
1
\end{bmatrix}
$}
\end{align}
. Thus we can calculate the velocities after $2k$th collision for any non-negative integer $k$. all terms letters initial velocities rely on only masses of each particle thus we can calculate for any initial condition.\\
When total collision number is $2k+1$, by applying $D_{AB}$ to $D^k$ from left side, we get velocities as
\begin{align}
&\begin{bmatrix}
v_{A2k+1}\\
v_{B2k+1}\\
v_{C2k+1}
\end{bmatrix}\\
&=\frac{1}{dM_{AB}}
\resizebox{\textwidth}{!}{$
\begin{bmatrix}
\left(m_A-3m_B\right)\left\{\left[\left(\overline{v_{AB}}-v_{C0}\right)\nu^2+c\frac{v_{B0}-v_{A0}}{2}\right]\sin{k\theta}-d\frac{v_{B0}-v_{A0}}{2}\cos{k\theta}\right\}\\
\left(m_B-3m_A\right)\left\{\left[\left(\overline{v_{AB}}-v_{C0}\right)\nu^2+c\frac{v_{B0}-v_{A0}}{2}\right]\sin{k\theta}-d\frac{v_{B0}-v_{A0}}{2}\cos{k\theta}\right\}\\
M_{AB}\left\{\left[\frac{v_{A0}-v_{B0}}{2}-c\left(\overline{v_{AB}}-v_{C0}\right)\right]\sin{k\theta}-d\left(\overline{v_{AB}}-v_{C0}\right)\cos{k\theta}\right\}
\end{bmatrix}
+\overline{v_{AB}}
\begin{bmatrix}
1\\
1\\
1
\end{bmatrix}
$}
\end{align}
Hence we can calculate velocities of each particle after any number of collisions with any initial condition. Now, the determinant of $D$ is
\begin{align}
\det{D}=\lambda_1\lambda_+\lambda_-=1
\end{align}.
For general, determinant being not zero means that the matrix is invertible, which is equivalent to existing inverse. This allows us to calculate initial velocities from velocity vector after collisions by applying inverse of $D^k$ or $D_{AB}D^k$. In contrast, if the determinant is zero, the matrix is not invertible, meaning, in this case, time reversal symmetry is lost. However, it is not realized this time thus that never happens.
\section{Maximum Number of Collisions}
To think about the number of collisions, we consider vector $\bm{u}$ below
\begin{align}
\bm{u}=
\begin{bmatrix}
\sqrt{m_A}v_A\\
\sqrt{m_B}v_B\\
\sqrt{m_C}v_C
\end{bmatrix}
\end{align}
. With this, total kinetic energy of whole system $K$ can be simplified as
\begin{align}
K=\frac12||\bm{u}||^2
\end{align}
. Now define a vector $\bm{n_{AB}}$ which takes relative velocity of A  and B from $\bm{u}$ as
\begin{align}
\bm{n_{AB}}=(\frac{1}{\sqrt{m_A}},-\frac{1}{\sqrt{m_B}},0)\\
\therefore \bm{n_{AB}}\cdot\bm{u}=v_A-v_B
\end{align}
. Similarly,
\begin{align}
\bm{n_{BC}}=(0,\frac{1}{\sqrt{m_B}},-\frac{1}{\sqrt{m_C}})
\end{align}
is defined as a vector which takes relative velocity of B and C from $\bm{u}$. According this, the collide condition can be replaced to
\begin{align}
n=2k:\bm{n_{BC}}\cdot\bm{u}<0\\
n=2k+1:\bm{n_{AB}}\cdot\bm{u}<0
\end{align}
. Moreover, we introduce vector $\bm{s}$ which is 
\begin{align}
\bm{s}=
\begin{bmatrix}
\sqrt{m_A}\\
\sqrt{m_B}\\
\sqrt{m_C}
\end{bmatrix}
\end{align}
then calculate the inner product of it and $\bm{n_{AB}}$ and $\bm{n_{BC}}$ which results
\begin{align}
\bm{n_{AB}}\cdot\bm{s}=\bm{n_{BC}}\cdot\bm{s}=0
\end{align}
 therefore we see that $\bm{s}$ and $\bm{n_{AB}}$ are orthogonal and $\bm{n_{AB}}$ and $\bm{n_{BC}}$ exist on a plane which is orthogonal with $\bm{s}$. Let the plane be $s^\perp$. Now we focus on only the relative velocity so we see the behavior of $\bm{u}$ in $s^\perp$. Thus, let $\bm{u}_\perp$ denote the component of $\bm{u}$ in $s^\perp$ that leads 
\begin{align}
\bm{u}_\perp=\bm{u}-(\bm{u}\cdot\hat{\bm{s}})\hat{\bm{s}}
\end{align}
. Here let $\hat{\bm{s}}$ be a unit vector directing same direction with $\bm{s}$. At the time, the $\bm{u}$ after collision of A and B, $\bm{u}'$, leads the relationship as 
\begin{align}
\bm{u}'=u-2\frac{\bm{u}\cdot\bm{n_{AB}}}{||\bm{n_{AB}}||^2}\bm{n_{AB}}
\end{align}
This is a transformation known as Householder transformation and hereafter we make consideration based on it. Similarly, after the collision between B and C, the below is led.
\begin{align}
\bm{u}'=u-2\frac{\bm{u}\cdot\bm{n_{BC}}}{||\bm{n_{BC}}||^2}\bm{n_{BC}}
\end{align}
At the collision of A and B, taking the unit vectors along
\begin{align}
\bm{s_{AB}}=
\begin{bmatrix}
\sqrt{m_A}\\
\sqrt{m_B}
\end{bmatrix}\\
\bm{n_{AB}}=
\begin{bmatrix}
\frac{1}{\sqrt{m_A}}\\
-\frac{1}{\sqrt{m_B}}
\end{bmatrix}
\end{align}
as bases to think $\bm{u}$, we can say
\begin{align}
\bm{u}=a\hat{\bm{s}}+b\hat{\bm{n}}_{AB}
\end{align}
because they are orthogonal. Solving it for $a,b$, 
\begin{align}
a=\frac{m_Av_A+m_Bv_B}{||\bm{s_{AB}}||}\\
b=\frac{v_A-v_B}{||\bm{n_{AB}}||}
\end{align}
When the collision occurs the sign of relative velocity is inversed and total momentum is conserved, which means $a$ is constant and sign of $b$ is inversed. That leads
\begin{align}
\bm{u}'=a\hat{\bm{s}}-b\hat{\bm{n}}_{AB}
\end{align}
This corresponds to mirror reflection of $\bm{n}_{AB}$ with constant line on cartesian coordinate, known as Householder reflection. Let $L_{AB}$ and $L_{BC}$ be the reflection lines in $s^\perp$. The wedge angle is $\Omega=\arccos(\hat{\bm{n}}_{AB}\!\cdot\!\hat{\bm{n}}_{BC})\in(0,\pi)$,
and the wedge set is 
\begin{align}
\mathcal{W}=\{\bm{u}\in s^\perp \mid \bm{u}\!\cdot \bm{n}_{AB}<0,\ \bm{u}\!\cdot \bm{n}_{BC}<0\,\}.
\end{align}
Each $AB$ (resp.\ $BC$) collision reflects $u_\perp$ across $L_{AB}$ (resp.\ $L_{BC}$).
Defining one cycle as $AB\to BC$, applying $D=D_{BC}D_{AB}$ to the initial state advances
the phase by \(\theta=\theta_{AB}+\theta_{BC}\), and the number of completed cycles is
\(K_{\max}=\big\lfloor(\Omega-\phi_{BC})/\theta\big\rfloor\).
where $\phi_{BC}$ is a phase of $\bm{u}$ after first collision of A and B measured from $L_{BC}$. To calculate this, we introduce a unit vector along a new axis which is orthogonal with both $\hat{\bm{n}}_{BC}$ and $\hat{\bm{s}}$ as
\begin{align}
\hat{\bm{\ell}}_{BC}=\frac{\hat{\bm{s}}\times\hat{\bm{n}}_{BC}}{||\hat{\bm{s}}\times\hat{\bm{n}}_{BC}||}
\end{align}
According to (3.9) and property of cross product of vectors, all these vectors are orthogonal with each other. So, let  $\phi_{BC}$ be an angle between $\bm{u}_\perp^{(1)}$, which is $\bm{u}_\perp$ after first collision between A and B, and $\hat{\bm{n}}_{BC}$ on a plane constituted from $\hat{\bm{\ell}}_{BC}$ and $\hat{\bm{n}}_{BC}$
\begin{align}
  \phi_{BC} &=
  \mathop{\atanTwo}\!\Big(
    \boldsymbol u^{(1)}_{\perp}\!\cdot\!\hat{\boldsymbol n}_{BC}\,,\;
    \boldsymbol u^{(1)}_{\perp}\!\cdot\!\hat{\boldsymbol\ell}_{BC}
  \Big)
\end{align}
\pgfmathsetmacro{\OmegaDeg}{60}
\pgfmathsetmacro{\phiDeg}{25}
\pgfmathsetmacro{\rOmega}{1.20}
\pgfmathsetmacro{\rphi}{0.95}
\pgfmathsetmacro{\ru}{1.45}

\begin{figure}[t]
\centering
\begin{tikzpicture}[scale=2,>=Stealth]
  \draw[->] (-0.2,0) -- (2.1,0) node[below right,yshift=-3pt] {$\hat{\boldsymbol\ell}_{BC}$};
  \draw[->] (0,-0.2) -- (0,2.1) node[above left,xshift=-2pt] {$\hat{\boldsymbol n}_{BC}$};

  \draw[very thick,gray] (0,0)--(2.0,0);
  \node[gray] at (2.0,0.36) {$L_{BC}$}; 

  \draw[very thick,gray,rotate=\OmegaDeg] (-0.2,0)--(2.0,0)
    node[above right] {$L_{AB}$};

  \draw (0,0) circle (1.6);

  \draw[->,thick] (\rOmega,0) arc[start angle=0, end angle=\OmegaDeg, radius=\rOmega];
  \node at ({\rOmega*cos(\OmegaDeg/2)}, {\rOmega*sin(\OmegaDeg/2)+0.16}) {$\Omega$};

  \draw[->,blue,thick] (\rphi,0) arc[start angle=0, end angle=\phiDeg, radius=\rphi];
  \node[blue] at ({0.80*cos(\phiDeg/2)}, {0.80*sin(\phiDeg/2)+0.06}) {$\phi_{BC}$}; 

  \draw[->,blue,thick] (0,0) -- ({\ru*cos(\phiDeg)}, {\ru*sin(\phiDeg)})
      node[pos=0.82, xshift=3pt, right, blue] {$\boldsymbol{u}^{(1)}_{\perp}$}; 
\end{tikzpicture}
\caption{Wedge in $s^\perp$ showing $L_{BC}$, $L_{AB}$, $\Omega$, and $\phi_{BC}$.}
\label{fig:wedge}
\end{figure}
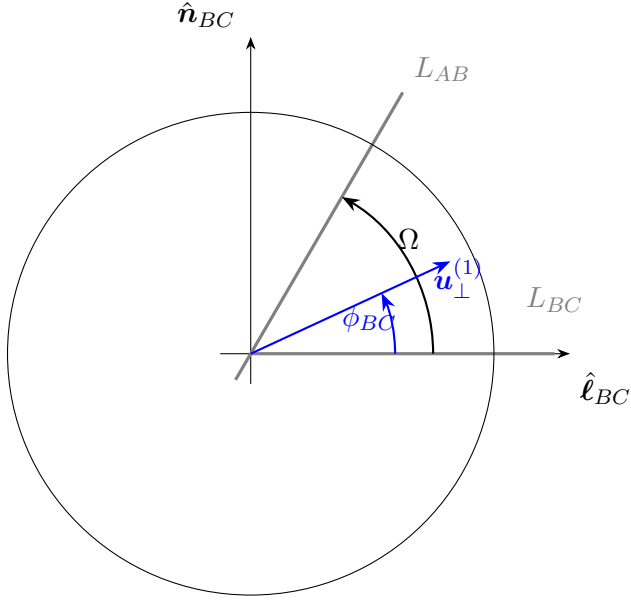

Based on this, we calculate maximum number of collisions  $n_{max}$. In a cycle, B collides twice and when $(\Omega-\phi_{BC})/\theta$ is not an integer number, there are still more phase to reflect after $K_{\mathrm{max}}$ cycles. Thus, another collision occurs. Moreover, B collide with A at first so adding it, total number of collisions in the system is

\begin{align}
n_{\mathrm{max}}=1+\left\lfloor \frac{\Omega-\phi_{BC}}{\theta}\right\rfloor+\left\lceil \frac{\Omega-\phi_{BC}}{\theta}\right\rceil
\end{align}
\section{Energy Transfer}
We consider the energy transfer between particles at collisions. The system is in vacuum, and only impulsive forces act at impacts; hence the total kinetic energy is conserved.
The change in A's energy across $2k+1$-th collision is
\begin{align}
\Delta E_{A2k+1}&=\frac{m_A}{2}\left(v_{A2k+1}^2-v_{A2k}^2\right)
\end{align}
. Now defining
\begin{align}
v_{A2k}=v_k+\overline{v_{AB}}\\
v_{B2k}=-v_k+\overline{v_{AB}}\\
v_{C2k}=v_k'+\overline{v_{AB}}
\end{align}
leads to 
\begin{align}
v_{A2k+1}=\frac{m_A-3m_B}{M_{AB}}v_k+\overline{v_{AB}}
\end{align}
which results
\begin{align}
\Delta E_{A2k+1}&=-\frac{4m_Am_Bv_k}{M_{AB}^2}P_{AB2k}
\end{align}

where $P_{AB2k}$ is the sum of momentum of A and B before the collision. At the time, the energy one lost is absorbed by another. Thus,
\begin{align}
\Delta E_{B2k+1}&=\frac{4m_Am_Bv_k}{M_{AB}^2}P_{AB2k}
\end{align}
.Similarly, the change of energy of C and B after $2k$th collision is calculated as
\begin{align}
\Delta E_{C2k}&=-\frac{4m_Bm_C}{M_{BC}^2}\frac{v_{k-1}+v_{k-1}'}{2}P_{BC2k-1}\\
\Delta E_{B2k}&=\frac{4m_Bm_C}{M_{BC}^2}\frac{v_{k-1}+v_{k-1}'}{2}P_{BC2k-1}
\end{align}
We consider that energy transfer is proportional to total momentum and average velocity based on average velocity of initial ones of A and B of two particles before collision. as the number of collisions increases, velocities are homogenised. Thus, the amount of energy changing through the collision must decrease.
\section{CONCLUSION}
We calculated the velocities of three particles in one dimension in vacuum with any mass ratio and initial condition through conservation of momentum formulated with matrices. Then, we evaluated the maximum number of collisions on a plane $s^\perp$. with $\bm{u}$, which expresses velocities , wedge between $L_{AB}$ and $L_{BC}$. It was resulted as 
\[
n=1+\big\lfloor(\Omega-\phi_{BC})/\theta\big\rfloor+\big\lceil(\Omega-\phi_{BC})/\theta\big\rceil,
\]
After that, we calculated the energy transfer. All of them were determined by initial condition and mass ratio. That strengthen the correctness of time reversal symmetry which is introduced in Newtonian mechanics.
\medskip

\appendix
\section*{Appendix A. SymPy script used to compute eigenvalues}

\begin{lstlisting}[language=Python, style=code,
  caption={Python/SymPy code to compute eigenvalues and eigenvectors of $M$},
  label={lst:sympy-eigs}]
import sympy as sp

mA, mB, mC = sp.symbols('mA mB mC', positive=True, real=True)
MAB = mA + mB
MBC = mB + mC
MABC = mA + mB + mC

M = (1 / (MAB * MBC)) * sp.Matrix([
    [MBC*(mA - mB),       2*MBC*mB,                    0],
    [2*mA*(mB - mC),      (mB - mA)*(mB - mC),         2*MAB*mC],
    [4*mA*mB,             2*mB*(mB - mA),              MAB*(mC - mB)]
])

eigen_data = M.eigenvects()

for i, (eigval, mult, eigvecs) in enumerate(eigen_data):
    lam = sp.Symbol(f'\\lambda_{i+1}')
    vec = eigvecs[0]

    print(f"eigenvalue \\( \\lambda_{i+1} = {sp.latex(eigval)} \\)")

    print(f"eigenvector \\( \\boldsymbol{{v}}_{i+1} = {sp.latex(vec)} \\)")

    
    lhs = M * vec
    rhs = eigval * vec
    residual = sp.simplify(lhs - rhs)

    print(f"\\( M \\boldsymbol{{v}}_{i+1} - \\lambda_{i+1} \\boldsymbol{{v}}_{i+1} = {sp.latex(residual)} \\)\\n")
\end{lstlisting}

\end{document}